\documentclass[reprint,aps,prl,floatfix]{revtex4-1} 
\usepackage{newtxtext}
\usepackage{newtxmath}
\usepackage{graphicx}
\usepackage{bm}
\usepackage{physics}
\usepackage{mathtools,amsfonts,dsfont}
\usepackage{subdepth}
\usepackage[unicode=true, pdfusetitle, bookmarks=true, bookmarksnumbered=false, bookmarksopen=false, breaklinks=true, pdfborder={0 0 0}, backref=false, colorlinks=true, linkcolor=blue, citecolor=blue, urlcolor=blue]{hyperref}
\allowdisplaybreaks
\newcommand{\sm}{\kern0.1em}

\begin{document}

\title{Exact propagating Dirac wave packets in an attractive Coulomb-like potential}

\begin{abstract}
We construct exact, positive-energy, normalizable wave-packet solutions of the Dirac equation in the axisymmetric potential \(\smash{V=-\,v_0/\rho}\)---to our knowledge, the first such solutions in any external potential. Remarkably, one family comprises only elementary functions whose longitudinal profiles reproduce the free-Schr\"odinger Hermite--Gauss wave packets in the nonrelativistic limit. All packets share two striking features: (i) a probability density that is pointwise decoupled from spin orientation---despite the inherent spin-orbit coupling of the Dirac equation---and (ii) a complete freezing of their time evolution at the critical coupling \(\smash{v_0\to\hbar\sm c/2}\). We also present a simple scheme that maps solutions of the 2D Helmholtz equation to further exact Dirac wave packets.
\end{abstract}

\author{Siddhant Das}
\email{Siddhant.Das@physik.uni-muenchen.de}
\affiliation{Arnold Sommerfeld Center for Theoretical Physics, Fakult\"at f\"ur Physik, Ludwig-Maximilians-Universit\"at M\"unchen, Theresienstr.~37, D-80333 M\"unchen, Germany}
\affiliation{Mathematisches Institut, Ludwig-Maximilians-Universit\"at M\"unchen, Theresienstr.~39, D-80333 M\"unchen, Germany}
\affiliation{John Bell Institute for the Foundations of Physics, New York, NY 10003, United States}
\date{\today}

\maketitle

Despite a century of intensive study, the catalog of exact solutions of the Dirac equation consists almost entirely of stationary states; normalizable propagating wave-packet solutions in external potentials remain conspicuously absent (see \cite{das2021,BG} for a survey). The most notable recent additions are the helical electron beams of Bia{\l}ynicki-Birula and Bia{\l}ynicka-Birula in a uniform magnetic field \cite{Birula2023} (transversely localized only) and the nonspreading wave packets of Campos \emph{et al.}\ in a plane-wave laser field \cite{Campos} (longitudinally localized only). In this Letter, we construct the first exact, normalizable propagating wave-packet solutions for the attractive Coulomb-like potential \(\smash{V=-\,v_0/\rho}\)---among smooth axisymmetric electrostatic potentials, the only one for which the stationary Dirac equation is known to be solvable \cite{ViSh}. The stationary states closely parallel their spherical Coulomb counterparts \cite[Sec.~9.6]{GreinerWaveEquations} and are well known from channeling radiation, where relativistic electrons traversing aligned crystals experience an effective \(1/\rho\) confinement from the atomic strings \cite{F11,paraxial}.

Our search for these propagating solutions is further motivated by a concrete experimental proposal: in \cite{DD,Exotic}, we showed that a nonrelativistic (NR) Bohmian spin-$\tfrac{1}{2}$ particle traversing a cylindrical waveguide displays an arrival-time distribution markedly dependent on its spin orientation relative to the waveguide axis, with the transverse case exhibiting a sharp upper cutoff $\tau_{\max}$. This prediction remains the subject of active discussion in the literature \cite{Mike,GTZ1,*GTZ2,*SScomment,Drezet,Poirier2024,UmbraldDrezet,JozaniTumulka,WillPL}; see \cite{SciAm,Meier2025} for popular accounts.

In \cite{das2021}, we initiated a relativistic extension by constructing exact Dirac wave packets in a hard-walled cylindrical waveguide, which exhibit robust, spin-dependent quantum backflow \cite{Backflow,BFrelativistic,BirulaBF}---a prerequisite for the ``exotic'' arrival-time distributions of \cite{DD,Exotic}. A hard wall, however, forces an arbitrary choice of boundary condition for the first-order Dirac equation \footnote{Several possibilities exist---for instance, the MIT bag, the chiral bag, or the boundary condition \((\mathds{1} + \beta)\Psi = 0\) adopted in \cite{das2021}.}, whereas the attractive \(1/\rho\) potential confines the wave function naturally, requiring no such ad hoc prescriptions.

Although originating as a spin-off of the arrival-time problem---where analytical control over wave-packet propagation across large space-time domains is imperative---our results are of much broader interest. Physically, the wave packets exhibit an unexpected pointwise decoupling of the position-space density from spin polarization, and a complete temporal freezing at the critical coupling. Methodologically, we introduce a simple `H\(\to\)D' scheme that maps solutions of the 2D Helmholtz equation into Dirac wave packets in the $1/\rho$ potential. Being exact, they also provide benchmarks for numerical Dirac solvers, which must contend with difficulties such as fermion doubling \cite{Fermiondoubling}.

\textit{Setup and notation.} In cylindrical-polar coordinates \(\vb{r}\equiv(\rho,\phi,z)\), the Dirac equation in an attractive \(1/\rho\) potential of coupling strength \(\smash{v_0 \,(> 0)}\) is given by
\begin{equation}\label{Deq}
    i\frac{\partial\Psi}{\partial t}
    = \hat{H}\sm\Psi=\left(\kern-0.1em\beta-i\bm{\alpha}\kern-0.15em\cdot\kern-0.16em\bm{\nabla} - \frac{v_0}{\rho}\!\right)\kern-0.1em\Psi,
\end{equation}
setting \(\smash{\hbar = m = c = 1}\) for convenience \cite{hmc}. We use the standard Dirac--Pauli representation for the Dirac matrices (\(\beta\) diagonal). As with the spherical Coulomb potential \cite[Sec.~9.6]{GreinerWaveEquations}, the coupling constant is bounded, \(\smash{v_0<1/2}\) (corresponding to \(\simeq98.663\text{ eV}\!\cdot\text{nm}\)), ensuring that \(\hat{H}\) admits a distinguished self-adjoint realization \cite{Gallone,DES} and, consequently, that the time evolution is unitary.

\textit{Stationary states.} Our building blocks are a class of odd-parity, degenerate eigenstates of \(\hat{H}\) labeled by the longitudinal wave number \(k\): \(\ket{\uparrow,k}\) and \(\ket{\downarrow,k}\) \cite{updown}, which correspond to the \emph{positive} energy 
\begin{equation}\label{energy}
    E_k = \cos\zeta\sqrt{1+k^2}.
\end{equation}
Here, \(\smash{\zeta=\sin^{-1}(2\sm v_0)}\), so that \(\zeta\in(0,\pi/2)\). Introducing the scalar envelope
\begin{equation}
    \mathcal{E}_k(\rho) =\rho^{(\cos\zeta\,-\,1)/2}\!\exp(-\,\rho\sin\zeta\sqrt{1+k^2}\,),
\end{equation}
which encodes both the radial decay and the square-integrable singularity at $\smash{\rho=0}$ (analogous to the spherical Coulomb ground states) common to both eigenstates, we have:
\begin{subequations}\label{omegakets}
\begin{align}
   \braket{\vb{r}\sm}{\uparrow,k}&=\mathcal{E}_k(\rho)\mqty(\displaystyle\frac{\cos\frac{\zeta}{2}}{\omega_-(k)}\sm\sin(kz) \label{Oup}\\[8pt]\displaystyle\frac{\sin\frac{\zeta}{2}}{\omega_+(k)}\cos(kz)\sm e^{i\phi}\\[7pt]\displaystyle-\,\frac{\cos\frac{\zeta}{2}}{\omega_+(k)}\sm i\cos(kz)\\[8pt]\displaystyle \frac{\sin\frac{\zeta}{2}}{\omega_-(k)}\sm i\sin(kz)\sm e^{i\phi}),\\[5pt]
    \braket{\vb{r}\sm}{\downarrow,k}&=\mathcal{E}_k(\rho)\mqty(\displaystyle-\,\frac{\sin\frac{\zeta}{2}}{\omega_+(k)}\sm\cos(kz)\sm e^{-\,i\phi}\\[8pt]\displaystyle\frac{\cos\frac{\zeta}{2}}{\omega_-(k)}\sin(kz)\\[7pt]\displaystyle\frac{\sin\frac{\zeta}{2}}{\omega_-(k)}\sm i\sin(kz)\sm e^{-\,i\phi}\\[8pt]\displaystyle \frac{\cos\frac{\zeta}{2}}{\omega_+(k)}\sm i\cos(kz)), \label{Odn}
\end{align}
\end{subequations}
where \[\omega_\pm(k)=k^{-1}\sqrt{\sqrt{k^2+1}\pm 1}.\] (See Supplemental Material (SM) for their explicit construction.) For all \(\smash{k,\sm k^\prime>0}\) and \(\smash{s,\sm s^\prime\in\{\uparrow,\downarrow\}}\), the continuum orthonormalization condition 
\begin{equation}\label{balanced}
    \braket{\sm s,k}{\sm s^\prime\!,k^\prime\sm} = \Lambda(\zeta)\sm\big(1+k^2\big)^{-\sm\cos\zeta/2}\,  \delta\big(k-k^\prime\sm\big)\sm \delta_{s\sm s^\prime}
\end{equation}
holds, with \(\smash{\Lambda(\zeta)=(2\sm\pi)^2\Gamma(1+\cos\zeta)\sm(2\sin\zeta)^{-1-\cos\zeta}}\).

\textit{Propagating wave packets.} We construct normalizable wave packets by superposing these basis states with suitable weighting functions \(a(k)\):
\begin{equation}\label{superpose}
    \ket{s;t} = \sqrt{\frac{2}{\pi}} \int_0^{\infty} \!\!\! dk ~ a(k)\sm e^{-\,it E_k} \ket{s,k}.
\end{equation}
Because this superposition involves only positive-energy basis states \(\smash{(E_k>0)}\), the resulting wave packets are free from Zitterbewegung (see also \cite{BBComment}). For these wave packets to be properly normalized, the weights must satisfy:
\begin{align}\label{NNorm}
    1 = \braket{s;t}{s;t} \overset{\eqref{balanced}}{=} 2\sm \Lambda(\zeta)\! \int_0^{\infty}\!\frac{dk}{\pi} \, \frac{|a(k)|^2}{\big(1 + k^2\big)^{\cos\zeta/2}}.
\end{align}
Consequently, since \(\smash{\cos\zeta>0}\), any locally square-integrable weight with \(|a(k)|^{2}=\mathcal{O}(1/k)\) as \(k\to\infty\) yields a normalizable wave packet \eqref{superpose}.

Introducing the pair of integrals
\begin{equation}\label{IPM}
    I_{\pm}(z,w)=\sqrt{\frac{2}{\pi}}\int_0^{\infty}\!\!\!dk~\frac{a(k)}{\omega_\pm(k)}\,\mqty{\displaystyle\cos\\[-3pt]\displaystyle\sin}\sm(kz)\sm e^{-\,w\sqrt{1+k^2}},
\end{equation}
defined for all \(\smash{z\in\mathbb{R}}\) and \(\smash{w\in\mathbb{C}}\) with \(\smash{\Re w>0}\), we can represent \eqref{superpose} in position space as follows:
\begin{subequations}\label{both}
    \begin{align}
    \Psi_{\uparrow}(\vb{r},t)&=\rho^{(\cos\zeta-1)/2}\sm\mqty(\displaystyle\cos\frac{\zeta}{2}\, I_-\\[5pt]\displaystyle \sin\frac{\zeta}{2}\, I_+\sm e^{i\phi}\\[5pt]\displaystyle-\,i\cos\frac{\zeta}{2}\, I_+\\[5pt]\displaystyle i\sin\frac{\zeta}{2}\, I_-\sm e^{i\phi}),\label{botha}\\
    \Psi_{\downarrow}(\vb{r},t) &=\rho^{(\cos\zeta-1)/2}\sm\mqty(\displaystyle-\,\sin\frac{\zeta}{2}\, I_+\sm e^{-\,i\phi}\\[5pt]\displaystyle \cos\frac{\zeta}{2}\, I_-\\[5pt]\displaystyle i\sin\frac{\zeta}{2}\, I_-\sm e^{-\,i\phi}\\[5pt]\displaystyle i\cos\frac{\zeta}{2}\, I_+),\label{bothb}
\end{align}
\end{subequations}
denoting \(\smash{\Psi_s=\braket{\vb{r}\sm}{\sm s;t}}\) and \(I_\pm = I_\pm(z,\rho\sin\zeta + it\cos\zeta)\) for brevity.

Substituting \eqref{both} into the Dirac equation \eqref{Deq} as an Ansatz (leaving \(I_\pm\) completely unspecified), the angular dependences, radial derivatives, and the singular \(\smash{1/\rho}\) potential all cancel, compressing the four coupled equations into a single pair of PDEs:
\begin{equation}\label{Ipde}
    \partial_zI_\pm \sm\mp\sm \partial_wI_\mp = I_\mp.
\end{equation}
Therefore, any pair of functions \(I_\pm(z,w)\) satisfying these PDEs yields an exact solution of Eq.~\eqref{Deq} via \eqref{both}. The integral representations \eqref{IPM} are one such pair, as may be verified directly using the algebraic properties of \(\omega_\pm\) [SM, Eq.~\eqref{iden}].

\begin{figure*}[!ht]
    \centering
    \includegraphics[width=\textwidth]{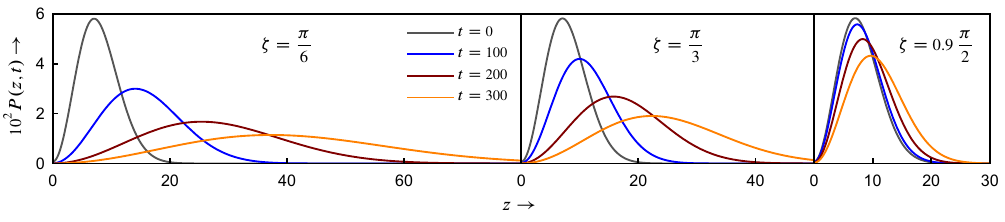}
    \caption{Time evolution of the marginal density \eqref{marginal} for the elementary-function wave packet with \(\smash{w_0=50}\). The propagation slows as the confinement \(\zeta\) strengthens, freezing completely as \(\smash{\zeta\to\pi/2}\).}\label{Fig1}
\end{figure*} 

\textit{Temporal freezing at critical coupling.} In these wave packets, time enters only through the complex combination \(\rho\sin\zeta + it\cos\zeta\), with \(\zeta\) interpolating between the free-motion limit \(\smash{(\zeta \to 0)}\), in which the dependence on \(\rho\) disappears, and the critical confinement limit \(\smash{(\zeta \to \pi/2)}\), where the \(t\)-dependence is lost and time evolution freezes. No such temporal freezing manifests in a NR treatment of the \(1/\rho\) potential \cite{2dhydrogen}, where \(v_0\) is unrestricted.

\textit{Probability density and spin-decoupling.} A short calculation using \eqref{both} shows that \(\smash{\Psi_\uparrow^{\dagger}\Psi_{\uparrow}^{\phantom{\uparrow}}=\Psi_{\downarrow}^{\dagger}\Psi_{\downarrow}^{\phantom{\uparrow}}}\) and \(\smash{\Psi_{\uparrow}^{\dagger}\Psi_{\downarrow}^{\phantom{\uparrow}}\!=0}\). It follows that the general superposition
\begin{equation}\label{superpose12}
\Psi=\cos(\vartheta/2)\sm \Psi_{\uparrow} \sm+\sm e^{i\varphi}\sin(\vartheta/2)\sm \Psi_{\downarrow},
\end{equation}
---which becomes spin-polarized along \(\smash{(\vartheta, \varphi)\in S^2}\) in the NR limit (see below)---has position-space density
\begin{equation}\label{density}
    \Psi^\dagger\Psi=\rho^{\cos\zeta-1}\Big(\sm\big|I_+\big|^2 + \big|I_-\big|^2\Big),
\end{equation}
independent of the spin-polarization angles \(\vartheta\) and \(\varphi\) at every spatial point---a hallmark of NR spin eigenstates. In relativistic quantum mechanics, however, spin and orbital angular momentum are not conserved separately (cf.~\cite{updown}), and the spin-orbit coupling inherent to the Dirac equation in an external potential makes such a cancellation atypical. Indeed, this cancellation fails for the hard-wall waveguide solutions of \cite{das2021}, as well as for the cylindrical step potential solutions of \cite{Leary2008}; in both, a generic spin superposition likewise carries a spin-orientation-dependent density---underscoring that the spin decoupling stems from the Coulomb--Dirac mode structure rather than from cylindrical symmetry \emph{per se}.

We now present one weight \(a(k)\) for which \(I_\pm\) evaluate in closed form; further closed-form solutions---in complementary error and Bessel functions---are collected in the End Matter (EM).

\textit{Explicit example 1.} Letting
\begin{equation}\label{simp}
    a(k)=N\sm\frac{k\sm e^{-\sm w_0\sqrt{1+k^2}}}{\sqrt{1+k^2}},
\end{equation}
in \eqref{IPM}, where \(\smash{w_0>0}\) controls the width of the wave packet and \(N\) is a normalization constant fixed by Eq.~\eqref{NNorm}, we obtain (see SM and EM for independent derivations)
\begin{equation}\label{IpmGold}
    I_\pm=N\sm\mqty(\text{sgn }z\\-\,1)\,\frac{\partial}{\partial z}\,\frac{\sqrt{\sqrt{\widetilde{w}^2+z^2} \mp \widetilde{w}}}{\sqrt{\widetilde{w}^2+z^2}} \exp(\!-\sm\sqrt{\widetilde{w}^2+z^2}\,),
\end{equation}
with \(\smash{\widetilde{w}=w_0+\rho\sin\zeta + it\cos\zeta}\).

Despite the singular Coulomb-like confinement, this fully relativistic solution comprises only elementary functions---mirroring the analytical simplicity of the free-Schr\"odinger Gaussian wave packet. Such an elementary form enables a direct symbolic check that the wave packets \eqref{both} satisfy the four coupled equations of \eqref{Deq}, as demonstrated in the Supplemental Mathematica notebook \cite{Mathematica}.

\textit{Nonrelativistic limit.} In the NR limit \(\smash{c\to\infty}\), the packet reduces longitudinally to the first excited free-Schr\"odinger Hermite--Gauss wave packet. Rescaling the dimensionless width \(w_0 = m\sm c^2 t_0/\hbar = \mathcal{O}\big(c^2\big)\) and restoring \(\hbar\), \(m\), and \(c\) via \cite{hmc}, the envelope exponent expands as
\begin{align}
    \sqrt{\widetilde{w}^2 + z^2} &\mapsto \frac{m\sm c^2}{\hbar}\sm (t_0+it) \,+\, \frac{it}{\hbar} E_\perp \,+\, k_\perp\sm\rho\nonumber\\
    &\qquad  +\frac{m}{2\sm\hbar}\kern-0.1em\left(\frac{z^2}{t_0+it}\right) \,+\, \mathcal{O}\big(c^{-2}\big),
\end{align}
where \(\smash{k_\perp = 2\sm m v_0/\hbar^2}\) and \(\smash{E_\perp = -\,\hbar^2 k_\perp^2/(2\sm m)}\). Each term has a clear origin. At leading order, the real part is neutralized by the normalization constant \(N\sim \sqrt{2}\,\zeta\sm(w_0/\pi)^{3/4}e^{w_0}\), while the imaginary component yields the standard rest-mass oscillation. At \(\mathcal{O}\big(c^0\big)\), the transverse \((\perp)\) terms furnish the known radial decay and binding-energy phase of the 2D Coulomb ground state \cite{2dhydrogen}, while the remaining longitudinal term yields a freely spreading Gaussian envelope.

The bispinor structure simplifies in tandem. Because \(\smash{\sin(\zeta/2)}\) and \(\smash{I_+/I_-}\) are both \(\mathcal{O}(1/c)\), only the \(\smash{\cos(\zeta/2)\sm I_-}\) components of \eqref{both} survive at \(\mathcal{O}\big(c^0\big)\). As a result, \(\Psi\) collapses onto a space-spin-factorized Pauli spinor with constant spin part \(\smash{\big(\kern-0.1em\cos(\vartheta/2),\, e^{i\varphi}\sin(\vartheta/2)\big)^{\!\top}}\!\) and a spatial part augmented by a \(\smash{(t_0+it)^{-\sm3/2}z}\) factor inherited from the odd-parity basis states \eqref{omegakets}---characteristic of the first-excited Hermite--Gauss wave packet.

\textit{Relativistic dispersion.} The static transverse profile found above persists into the relativistic regime unless \(\smash{w_0 \ll 1}\), so the longitudinal marginal density
\begin{equation}\label{marginal}
    P(z,t) = 2\sm\pi\!\int_0^{\infty}\!\!\!d\rho\,\rho~\Psi^\dagger\Psi
\end{equation}
suffices to characterize the evolution. By the odd parity of \(\Psi\), \(P\) is even in \(z\); therefore, we consider only \(\smash{z>0}\). Figure \ref{Fig1} shows the resulting time evolution: dispersive spreading is evident and slows markedly as the confinement strengthens.

This slowing has an optical interpretation: the \(1/\rho\) potential acts like a dielectric medium, with effective refractive index \(\smash{n=\sec\zeta}\), retarding the longitudinal group velocity 
\begin{equation}
   v_g=\partial_k E_k \overset{\eqref{energy}}{=} \frac{k\cos\zeta}{\sqrt{1+k^2}}<\cos\zeta,
\end{equation}
below its vacuum value \(\smash{(=\!1)}\). As \(\smash{\zeta\to\pi/2}\), the effective light speed---and with it the group velocity---vanishes, driving the temporal freezing noted above.

\begin{figure}[!ht]
    \centering
    \includegraphics[width=\columnwidth]{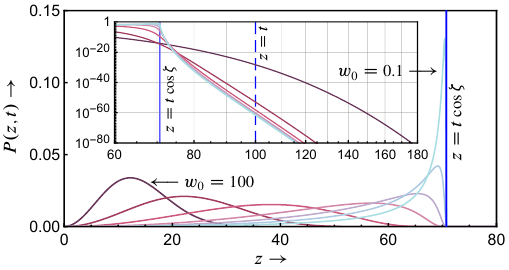}
    \caption{Marginal density \eqref{marginal} at \(\smash{t=100}\) and \(\smash{\zeta=\pi/4}\), for \(w_0\) from 100 (darkest) to \(0.1\) (lightest). As \(w_0\) falls below the Compton scale \(\smash{(\simeq1)}\), the leading edge sharpens against the effective light cone \(\smash{z=t\cos\zeta}\) (solid blue). Inset: log-log view of the tails reaching beyond both \(\smash{z=t\cos\zeta}\) and the true light cone \(\smash{z=t}\) (dashed blue).}\label{Fig2}
\end{figure}

\textit{Sub-Compton localization.} Having varied the coupling \(\zeta\) at fixed \(w_0\), we now fix \(\smash{\zeta<\pi/2}\) and vary the width parameter. For $\smash{w_0\gg1}$, the marginal density exhibits a Gaussian-like profile reflecting its NR limit. However, as \(w_0\) approaches the Compton scale, ultrarelativistic momentum components \(\smash{k\gg1}\) dominate \eqref{IPM}, severely skewing the profile (see Fig.~\ref{Fig2}). The packet piles up asymmetrically against the effective light cone \(\smash{z=t\cos\zeta}\) set by the group velocity. More strikingly, exponentially small tails persist even beyond the true light cone \(\smash{z=t}\) (Fig.~\ref{Fig2}, inset)---an explicit, closed-form manifestation of Hegerfeldt's localization theorem for positive-energy states \cite{Hegerfeldt}.

\textit{Schemes for generating additional exact solutions.} The explicit elementary-function wave packet is not an isolated special case; several systematic approaches yield further closed-form solutions of Eq.~\eqref{Deq}.

The simplest method is differentiation: because \eqref{Deq} is linear and \(\hat{H}\) commutes with $\partial_z$ and $\partial_{w_0}$, differentiating \eqref{IpmGold} to any order with respect to either variable yields new closed-form wave packets.

More powerfully, Eqs.~\eqref{Ipde} can be solved explicitly from a single scalar seed: any \(\Upsilon(z,w)\) obeying the Helmholtz equation (HE),
\begin{equation}\label{HE}
    \big(\partial_z^2+\partial_w^2\big)\Upsilon = \Upsilon,
\end{equation}
generates an exact Dirac wave packet \eqref{both} via
\begin{equation}\label{HEtoD}
    I_\pm(z,w) = p_\pm\sm\big(\Upsilon \pm \partial_w\!\Upsilon\big) \,+\, p_\mp\sm\partial_z\!\Upsilon,
\end{equation}
as verified by direct substitution into \eqref{Ipde}. The arbitrary complex constants \(p_\pm\) yield a two-parameter family per seed. Normalizability selects seeds with \(\smash{\Upsilon\to 0}\) as \(\smash{|z|\to\infty}\) and \(\smash{\Re w\to\infty}\) \footnote{Equation \eqref{Ipde} may be viewed as a mass-deformed analog of the Cauchy--Riemann system, with the HE \eqref{HE} in the role of Laplace's equation; the map \eqref{HEtoD} correspondingly reconstructs a holomorphic function from a single harmonic seed.}.

The symmetries of the HE further extend the map's reach by turning seeds into new seeds: translations \((z,w)\mapsto(z-z^\prime,w-w^\prime\sm)\), the reflection \(\smash{(z,w)\mapsto(w,z)}\), and the rotations \(\smash{(z,w)\mapsto R(\theta)\sm(z,w)}\) with \(\smash{w^\prime,\sm z^\prime,\sm \theta\in\mathbb{C}}\), etc., all preserve \eqref{HE}; a purely imaginary \(\theta\), in particular, acts as a hyperbolic squeeze, stretching the packet along one axis while compressing the other. Furthermore, since the HE separates in several 2D curvilinear coordinate systems, its canonical solutions---Bessel, parabolic-cylinder, Mathieu---project directly into corresponding families of exact Dirac wave packets (see the EM for the Bessel case).

In effect, the `H\(\to\)D' map provides a direct pipeline from the solution space of the 2D HE to that of Eq.~\eqref{Deq}. This construction parallels the `KG\(\to\)D' map of \cite{Birula2017} for the free Dirac equation, but with broader reach: where that map draws on the sparse set of closed-form \((3+1)\)D Klein--Gordon solutions, ours taps the extensive catalog of closed-form solutions of the 2D HE.

\textit{Summary and outlook.} We have constructed families of normalizable, propagating wave-packet solutions of the Dirac equation in the attractive \(1/\rho\) potential: one in elementary functions and the rest in known special functions (EM). To our knowledge, they are the first in any external potential. The position-space probability density of every wave packet is pointwise independent of spin orientation. Whether this reflects a conserved operator commuting with \(\hat{H}\) (perhaps in the spirit of the Johnson--Lippmann symmetry of the spherical Coulomb problem \cite{JL}) remains open.

A second structural hallmark is the complete freezing of temporal evolution at the critical coupling \(\smash{v_0\to\hbar\sm c/2}\) (the self-adjointness threshold of \(\hat{H}\)). This relativistic effect stems from the collapse of the mode energies \(\smash{E_k \propto \cos\zeta\to 0}\). Even the inverse-square potential \cite{Griffiths2006}, which shares the same self-adjointness-threshold structure, shows no such freezing: as its critical coupling is approached, time evolution and wave-packet spreading proceed unabated.

The present wave packets also exhibit the stable spin-dependent quantum backflow reported for the hard-wall geometry in \cite{das2021} (demonstration omitted for brevity). A natural next step is a fully relativistic treatment of the arrival-time problem of \cite{DD,Exotic}, built on the closed-form packets reported here and in \cite{das2021}. Preliminary analysis indicates that the pronounced spin dependence of the arrival-time distributions survives in both smooth and hard-wall geometries and persists deep into the relativistic regime, beyond the reach of earlier NR treatments.

\begin{acknowledgments}
The author thanks Markus N\"oth and Franck Lalo\"e for helpful feedback on the manuscript.
\end{acknowledgments}

\bibliography{RefDCWG}

@book{GreinerWaveEquations,
  title={Relativistic Quantum Mechanics--Wave Equations},
  edition={third},
  author={Greiner, W.},
  year={2000},
  publisher={Springer-Verlag},
  address={Berlin Heidelberg},
  series={Texts and Monographs in Physics}
}

@book{GH,
  author = {Gradshteyn, I. S. and Ryzhik, I. M.},
  edition = {seventh},
  publisher = {Elsevier},
  address   = {New York},
  title = {Table of integrals, series, and products},
  year = {2007},
}

@incollection{Gallone,
  author = {Gallone, M.}, title = {Self-Adjoint Extensions of {D}irac Operator with {C}oulomb Potential},
  booktitle = {Advances in Quantum Mechanics}, editor = {Michelangeli, A. and Dell'Antonio, G.},
  series = {Springer INdAM Series}, volume = {18}, pages = {169--185},
  publisher = {Springer}, address = {Cham}, year = {2017}, doi = {10.1007/978-3-319-58904-6_10}}

@article{DES,
  author = {Dolbeault, J. and Esteban, M. J. and S\'er\'e, \'E.},
  title = {Distinguished self-adjoint extension and eigenvalues of operators with gaps. {A}pplication to {D}irac--{C}oulomb operators},
  journal = {J. Spectr. Theory}, volume = {13}, pages = {491--524}, year = {2023}, doi = {10.4171/JST/461}}

@article{Andrews2008,
  title = {The evolution of free wave packets},
  author = {Andrews, Mark},
  journal = {Am. J. Phys.},
  volume = {76},
  number = {12},
  pages = {1102--1107},
  year = {2008},
  doi = {10.1119/1.2982628}
}

@article{Leary2008,
  title = {Self-spin-controlled rotation of spatial states of a {D}irac electron in a cylindrical potential via spin--orbit interaction},
  author = {Leary, C. C. and Reeb, D. and Raymer, M. G.},
  journal = {New J. Phys.},
  volume = {10},
  number = {10},
  pages = {103022},
  year = {2008},
  doi = {10.1088/1367-2630/10/10/103022}
}

@article{hydrogen,
  title = {Schr{\"o}dinger and Dirac equations for the hydrogen atom and Laguerre polynomials},
  author = {Mawhin, J. and Ronveaux, A.},
  journal = {Arch. Hist. Exact Sci.},
  volume = {64},
  pages = {429-460},
  year = {2010},
  doi = {10.1007/s00407-010-0060-3}
}

@book{Prudnikov,
  title      = {Integrals and Series. Volume 4: Direct Laplace Transforms},
  author     = {Prudnikov, Anatolii Platonovich and Brychkov, Yurii Aleksandrovich and Marichev, Oleg Igorevich},
  translator = {Queen, N. M.},
  year       = {1992},
  publisher  = {Gordon and Breach Science Publishers},
  address    = {New York},
  isbn       = {2-88124-682-6}
}

@article{Griffiths2006,
  author  = {Essin, Andrew M. and Griffiths, David J.},
  title   = {Quantum mechanics of the $1/x^2$ potential},
  journal = {Am. J. Phys.},
  volume  = {74},
  number  = {2},
  pages   = {109--117},
  year    = {2006},
  month   = {Feb},
  doi     = {10.1119/1.2165248}
}

@article{das2021,
      title={Relativistic electron wave packets featuring persistent quantum backflow}, 
      author={Siddhant Das},
      year={2021},
      eprint={2112.13180},
      archivePrefix={arXiv},
      primaryClass={quant-ph},
      url={https://arxiv.org/abs/2112.13180}, 
}

@article{Birula2017,
  title = {Relativistic Electron Wave Packets Carrying Angular Momentum},
  author = {Bia{\l}ynicki-Birula, Iwo and Bia{\l}ynicka-Birula, Zofia},
  journal = {Phys. Rev. Lett.},
  volume = {118},
  issue = {11},
  pages = {114801},
  year = {2017},
  month = {Mar},
  publisher = {American Physical Society},
  doi = {10.1103/PhysRevLett.118.114801}
}

@article{DD,
   author = "Das, S. and D{\"{u}}rr, D.",
   title = "Arrival Time Distributions of Spin-1/2 Particles",
   journal = {Sci. Rep.},
   year = {2019},
   volume = {9},
   pages = {2242},
   doi = {10.1038/s41598-018-38261-4},
}

@article{Exotic,
  title = {Exotic {B}ohmian arrival times of spin-1/2 particles: An analytical treatment},
  author = {Das, Siddhant and N\"oth, Markus and D\"urr, Detlef},
  journal = {Phys. Rev. A},
  volume = {99},
  issue = {5},
  pages = {052124},
  numpages = {13},
  year = {2019},
  month = {May},
  publisher = {American Physical Society},
  doi = {10.1103/PhysRevA.99.052124},
 }

@article{Backflow,
	doi = {10.1088/0305-4470/27/6/040},
	year = 1994,
	volume = {27},
	number = {6},
	pages = {2197--2211},
	author = {Bracken, A. J. and Melloy, G. F.},
	title = {Probability backflow and a new dimensionless quantum number},
	journal = {J. Phys. A: Math. Gen.},
}

@article{BFrelativistic,
  author = {Melloy, G. F. and Bracken, A. J.},
  title = {Probability Backflow for a Dirac Particle.},
  journal = {Found. Phys.},
  volume = {28},
  pages = {505-514},
  year = {1998},
  doi = {10.1023/A:1018724313788},
}

@article{Hegerfeldt,
  author  = {Hegerfeldt, Gerhard C.},
  title   = {Remark on Causality and Particle Localization},
  journal = {Phys. Rev. D},
  volume  = {10},
  pages   = {3320},
  year    = {1974},
  doi     = {10.1103/PhysRevD.10.3320},
}

@article{BirulaBF,
doi = {10.1088/1751-8121/ac65c1},
year = {2022},
month = {may},
publisher = {IOP Publishing},
volume = {55},
number = {25},
pages = {255702},
author = {Bia{\l}ynicki-Birula, Iwo and Bia{\l}ynicka-Birula, Zofia and Augustynowicz, Szymon},
title = {Backflow in relativistic wave equations},
journal = {J. Phys. A: Math. Theor.}
}

@article{SciAm,
  title = {Can We Gauge Quantum Time of Flight?},
  author = {Ananthaswamy, A.},
  journal = {Sci. Am.},
  volume = {326},
  issue = {1},
  year = {2022},
  url = {https://www.scientificamerican.com/article/this-simple-experiment-could-challenge-standard-quantum-theory/},
  note = {{G}erman translation: \emph{Stoppuhr f\"ur die Quantenwelt}, \href{https://www.spektrum.de/magazin/zeitmessung-stoppuhr-fuer-die-quantenwelt/2008774}{Spektrum der Wissenschaft (2022)}}
}

@Article{Drezet,
AUTHOR = {Drezet, Aur{\'{e}}lien},
TITLE = {Arrival Time and {B}ohmian Mechanics: It Is the Theory Which Decides What We Can Measure},
JOURNAL = {Symmetry},
VOLUME = {16},
YEAR = {2024},
NUMBER = {10},
ARTICLE-NUMBER = {1325},
ISSN = {2073-8994},
DOI = {10.3390/sym16101325}
}

@article{GTZ1,
author = {Goldstein, Sheldon and Tumulka, Roderich and Zangh{\`\i}, Nino}, title = {On the spin dependence of detection times and the nonmeasurability of arrival times}, journal = {Sci. Rep.}, year = {2024}, volume = {14}, number = {1}, pages = {3775}, doi = {10.1038/s41598-024-53777-8}, publisher = {Nature Portfolio} }

@article{GTZ2,
  author    = {Sheldon Goldstein and Roderich Tumulka and Nino Zangh{\`\i}},
  title     = {Arrival Times Versus Detection Times},
  journal   = {Found. Phys.},
  volume    = {54},
  number    = {5},
  pages     = {1--25},
  year      = {2024},
  month     = {Sep},
  publisher = {Springer US},
  doi       = {10.1007/s10701-024-00798-y}
}

@article{Mike,
  author    = {James M. Wilkes},
  title     = {The Pauli and L{\'e}vy-Leblond equations, and the spin current density},
  journal   = {Eur. J. Phys.},
  volume    = {41},
  number    = {3},
  pages     = {035402},
  year      = {2020},
  month     = {Mar},
  publisher = {IOP Publishing},
  doi       = {10.1088/1361-6404/ab7495}
}

@article{JL,
  title = {Relativistic Kepler Problem},
  author = {Johnson, M. H. and Lippmann, B. A.},
  journal = {Phys. Rev.},
  volume = {78},
  issue = {3},
  pages = {329--329},
  numpages = {0},
  year = {1950},
  month = {May},
  publisher = {American Physical Society},
  doi = {10.1103/PhysRev.78.329}
}

@article{Birula2023,
doi = {10.1088/1751-8121/acdcd1},
url = {https://doi.org/10.1088/1751-8121/acdcd1},
year = {2023},
month = {jun},
publisher = {IOP Publishing},
volume = {56},
number = {28},
pages = {285302},
author = {Bia{\l}ynicki-Birula, Iwo and Bia{\l}ynicka-Birula, Zofia},
title = {Helical beams of electrons in a magnetic field: new analytic solutions of the Schr{\"o"}dinger and Dirac equations},
journal = {J. Phys. A: Math. Theor.}
}

@article{SScomment,
      title={Comment on ``the Spin Dependence of Detection Times and the Nonmeasurability of Arrival Times''}, 
      author={Siddhant Das and Serj Aristarhov},
      year={2023},
      eprint={2312.01802},
      archivePrefix={arXiv},
      primaryClass={quant-ph},
      url={https://arxiv.org/abs/2312.01802}, 
}

@article{Poirier2024,
  author    = {Bill Poirier and Richard Lombardini},
  title     = {Dwell Times, Wavepacket Dynamics, and Quantum Trajectories for Particles with Spin 1/2},
  journal   = {Entropy},
  volume    = {26},
  number    = {4},
  pages     = {336},
  year      = {2024},
  month     = {Apr},
  publisher = {MDPI},
  doi       = {10.3390/e26040336}
}

@article{Campos,
  title = {Nonspreading relativistic electron wavepacket in a strong laser field},
  author = {Campos, Andre G. and Hatsagortsyan, Karen Z. and Keitel, Christoph H.},
  journal = {Phys. Rev. Res.},
  volume = {6},
  issue = {2},
  pages = {023040},
  numpages = {10},
  year = {2024},
  month = {Apr},
  publisher = {American Physical Society},
  doi = {10.1103/PhysRevResearch.6.023040},
  url = {https://link.aps.org/doi/10.1103/PhysRevResearch.6.023040}
}

@book{BG,
  title={The Dirac Equation and its Solutions},
  volume=4,
  author={Bagrov, V. G. and Gitman, D.},
  year={2014},
  address={Berlin},
  publisher={de Gruyter},
  series={Studies in Mathematical Physics}
}

@article{Fermiondoubling,
title = {Finite element formulation of the Dirac equation and the problem of fermion doubling},
journal = {Phys. Lett. A},
volume = {242},
number = {4},
pages = {245-250},
year = {1998},
issn = {0375-9601},
doi = {https://doi.org/10.1016/S0375-9601(98)00218-7},
url = {https://www.sciencedirect.com/science/article/pii/S0375960198002187},
author = {C. M{\"u}ller and N. Gr{\"u}n and W. Scheid},
}

@misc{hmc,
  note = {The physical constants may be restored with the following dimensionless substitutions: \(\smash{\rho\mapsto\rho/\lambdabar}\), \(\smash{z\mapsto z/\lambdabar}\), \(\smash{t\mapsto ct/\lambdabar}\), \(\smash{v_0\mapsto v_0/(\hbar\kern0.1em c)}\), and \(\smash{\Psi\mapsto\lambdabar^{3/2}\Psi}\), where \(\smash{\lambdabar=\hbar/(mc)}\) is the ``reduced'' Compton wavelength.}
}

@misc{updown,
note = {The \(\uparrow\), \(\downarrow\) labels are mnemonic: these states are \emph{not} eigenstates of \(\hat{S}_z\); instead, they are eigenstates of \(\hat{J}_z=\hat{L}_z+\hat{S}_z\).}
}

@article{BBComment,
  title = {Comment on ``Nondispersive analytical solutions to the Dirac equation''},
  author = {Bia{\l}ynicki-Birula, Iwo and Bia{\l}ynicka-Birula, Zofia},
  journal = {Phys. Rev. Res.},
  volume = {2},
  issue = {3},
  pages = {038001},
  numpages = {3},
  year = {2020},
  month = {Jul},
  publisher = {American Physical Society},
  doi = {10.1103/PhysRevResearch.2.038001}
}

@article{UmbraldDrezet,
      title={Arrival-time distributions as a probe of the preferred foliation in relativistic {B}ohmian mechanics}, 
      author={Arnaud Amblard and Aurelien Drezet},
      year={2026},
      eprint={2604.17507},
      archivePrefix={arXiv},
      primaryClass={quant-ph},
      url={https://arxiv.org/abs/2604.17507}
}

@article{Meier2025,
  author    = {Christian J. Meier},
  title     = {Quantenmechanik: Schr{\"o}dingers Katze und die Realit{\"a}t},
  journal   = {S{\"u}ddeutsche Zeitung},
  year      = {2025},
  month     = {Sep},
  day       = {19},
  url       = {https://www.sueddeutsche.de/wissen/quantenmechanik-realitaet-beobachtung-li.3291232}
}

@article{Leviatan,
  title = {Symmetries and Supersymmetries of the Dirac Hamiltonian with Axially Deformed Scalar and Vector Potentials},
  author = {Leviatan, A.},
  journal = {Phys. Rev. Lett.},
  volume = {103},
  issue = {4},
  pages = {042502},
  numpages = {4},
  year = {2009},
  month = {Jul},
  publisher = {American Physical Society},
  doi = {10.1103/PhysRevLett.103.042502}
}

@article{JozaniTumulka,
      title={Detection Time Distribution Predicted Using Absorbing Boundary Conditions and Imaginary Potentials}, 
      author={Alireza Jozani and Roderich Tumulka},
      year={2026},
      eprint={2603.22044},
      archivePrefix={arXiv},
      primaryClass={quant-ph},
      url={https://arxiv.org/abs/2603.22044}, 
}

@misc{WillPL,
      title={The empirical consequences of the no-signaling hypothesis}, 
      author={Will Cavendish},
      year={2026},
      note="forthcoming"
}

@misc{Mathematica,
  author = {Wolfram Research{,} Inc.},
  title = {Mathematica, {V}ersion 14.3},
  url = {https://www.wolfram.com/mathematica},
  note = {Champaign, IL, 2025}
}

@article{2dhydrogen,
author = {Zaslow, B.  and Zandler, Melvin E. },
title = {Two-Dimensional Analog to the Hydrogen Atom},
journal = {Am. J. Phys.},
volume = {35},
number = {12},
pages = {1118-1119},
year = {1967},
doi = {10.1119/1.1973790}
}

@article{paraxial,
  title = {Dirac states of relativistic electrons channeled in a crystal and high-energy channeling electron-positron pair production by photons},
  author = {Olsen, Haakon A. and Kunashenko, Yuri},
  journal = {Phys. Rev. A},
  volume = {56},
  issue = {1},
  pages = {527--537},
  numpages = {0},
  year = {1997},
  month = {Jul},
  publisher = {American Physical Society},
  doi = {10.1103/PhysRevA.56.527}
}

@article{ViSh,
author = {Shishkin, German V.  and Villalba, Victor M. },
title = {Dirac equation in external vector fields: New exact solutions},
journal = {J. Math. Phys.},
volume = {30},
number = {10},
pages = {2373-2381},
year = {1989},
doi = {10.1063/1.528567},
}

@article{F11,
title = {Exact solution of Dirac equation for axially channeled relativistic electrons},
journal = {{N}uovo {C}imento C},
volume = {34},
number = {4},
pages = {111-118},
year = {2011},
issn = {1826-9885},
doi = {10.1393/ncc/i2011-10965-y},
author = {Korotchenko, K. B. and Kunashenko, Yu. P.},
}

\section*{End matter}\label{EM}
Of further interest to specialists, we record two closed-form families that complement the elementary solution \eqref{IpmGold}: a complementary-error-function family from a second weight, and a Bessel-function family \eqref{seedpolar} that illustrates the `H\(\to\)D' construction and contains \eqref{IpmGold} as a special case.

\textit{Explicit example 2.} A pure exponential weight,
\begin{equation}\label{c2}
    a(k)=N e^{-\sm z_0\sm k},\qquad N,\sm \Re z_0>0,
\end{equation}
yields closed-form solutions \eqref{both} involving the complementary error function erfc; the integrals \eqref{IPM} evaluate to (see SM for details)
\begin{equation}\label{ItoJ}
    I_\pm = N_\pm\sm \Big[J_\pm\big(z_0-iz,w\big) \,\pm\, J_\pm\big(z_0+iz,w\big)\Big],
\end{equation}
where \(\smash{N_+=iN_-=N/2}\), \(\smash{w=\rho\sm \sin\zeta+it\cos\zeta}\), and
\begin{align}\label{Jpm}
    J_\pm(z,w) &= \mp\,\frac{\partial}{\partial w}\frac{z}{\sqrt{w^2\!-z^2}}\!\left[\frac{\text{erfc}(\eta_+)}{\eta_\mp}\exp(\sqrt{w^2-z^2}\,)  \right.\nonumber\\[5pt]
    &\kern1.6cm\left.-\,\frac{\text{erfc}(\eta_-)}{\eta_\pm}\exp(-\sqrt{w^2\!-z^2}\,)\right]\!,
\end{align}
\(\smash{\eta_\pm = \sqrt{w\pm\sqrt{w^2-z^2}}}\). Additional closed-form solutions follow from applying \(\partial_z\), \(\partial_{z_0}\), and \(\smash{\hat{R}_z=\beta\Sigma_z - i\gamma_5\partial_z}\) \cite{Leviatan} to \eqref{both}, each commuting with \(\hat{H}\) and therefore producing new closed forms from \eqref{ItoJ} by elementary operations \footnote{Here \(\hat{R}_z\) acts as an independent generator, unlike for \eqref{IpmGold}, where it coincides with \(\partial_{w_0}\).}.

\textit{Explicit example 3.} A particularly tractable one-parameter family of Helmholtz seeds is
\begin{equation}\label{seedpolar}
    \Upsilon_\nu(z,w) = \big(q/R\big)^{\nu}\kern-0.1em K_\nu(R),
\end{equation}
where \(q=\widetilde{w}-iz\), \(R=\sqrt{\widetilde{w}^2+z^2}\), \(\widetilde{w}=w+w_0\) \((w_0>0)\). These are built from the modified Bessel function \(K_\nu\) and solve the HE for all \(\smash{\nu\in\mathbb{R}}\). Because \(R^{\nu}\partial_R\big(R^{-\sm\nu}K_\nu(R)\big)=-\sm K_{\nu+1}(R)\), both \(\partial_z\!\Upsilon_\nu\) and \(\partial_w\!\Upsilon_\nu\) are linear combinations of \(\Upsilon_\nu\) and \(\Upsilon_{\nu+1}\); the map \eqref{HEtoD} then yields exact Dirac wave packets \eqref{both} with
\begin{align}
    \mqty(I_+ \\[4pt] I_-) =\Upsilon_\nu\sm \mqty(p_+ \\[4pt] p_-) \,-\, \frac{1}{q} \sm\mqty(p_+ & p_- \\[4pt] -p_- & p_+) \mqty(-\sm\nu & \widetilde{w}\\[4pt] \,i\nu & z) \mqty(\Upsilon_\nu \\[4pt] \Upsilon_{\nu+1}),
\end{align}
normalizable thanks to the exponential decay of \(K_\nu(R)\) as \(\smash{|R|\to\infty}\); the singularity at \(\smash{q=R=0}\) is avoided since \(w_0\allowbreak >0\).

The index \(\nu\) sets the analytic character of the packet. Half-integer \(\nu\) collapses the modified Bessel functions---and with them the Dirac wave packets---to elementary functions \cite[Sec.~8.468]{GH}. In particular, since \(\smash{K_{1/2}(R)=e^{-R}\!\sqrt{\pi/(2R)}}\), feeding \eqref{HEtoD} the symmetrized Helmholtz seed \(\Upsilon_{1/2}(z,w)+\Upsilon_{1/2}(-\sm z,w)\) reproduces \eqref{IpmGold} for \(p_+=-\,N/\sqrt{\pi}\) and \(p_-=0\). Indeed, higher half-integer values of \(\nu\) systematically recover the higher-order Hermite--Gauss wave packets \cite{Andrews2008} in the NR limit, complementing the NR Laguerre--Gauss limit of the exponential free-Dirac beams derived in \cite{Birula2017}.

\newpage 

\onecolumngrid

\section*{Supplemental Material}
This Supplemental Material details the construction of the stationary basis states \eqref{omegakets} utilized in the main text. It also outlines the step-by-step evaluation of the longitudinal momentum integrals $I_\pm$ for both elementary- and complementary-error-function wave packets.
\subsection{Stationary basis states}\label{stationary states}
We seek stationary solutions of \eqref{Deq} of the form \(\smash{\Psi=e^{-\,iEt}\Phi(\rho,\phi,z)}\), with \(\Phi\) satisfying the time-independent Dirac equation (TIDE): \(\smash{\hat{H}\Phi=E\Phi}\), \(\smash{\hat{H}=\beta - i\bm{\alpha}\cdot\bm{\nabla} + V}\). Targeting the relativistic treatment of the arrival-time problem of \cite{DD,Exotic}, we restrict attention to positive-energy eigenstates \(\smash{(E>0)}\) with the simplest admissible radial and angular dependences. For a more comprehensive treatment of the stationary states, see \cite{F11}.

Symmetry considerations for translationally invariant, axisymmetric Dirac Hamiltonians \cite{Leviatan,F11} dictate the form:
\begin{equation}\label{Phi}
    \Phi(\rho,\phi,z) = e^{ikz\sm +\sm im_j\phi}\mqty(F(\rho)\sm e^{-\,i\phi/2}\\\frac{\mu-1}{k}\sm G(\rho)\sm e^{i\phi/2} \\ \frac{\mu-1}{k}\sm F(\rho)\sm e^{-\,i\phi/2}\\ G(\rho)\sm e^{i\phi/2}),
\end{equation}
where \(k\), \(m_j\), \(\mu\) denote the eigenvalues of the mutually commuting operators listed in Table \ref{EigenTab}.

\begin{table}[!ht]
\renewcommand{\arraystretch}{1.3}
\caption{The four mutually commuting operators of which \(\Phi\), Eq.~\eqref{Phi}, is a simultaneous eigenfunction, with their eigenvalues.}\vspace{1mm}
\label{EigenTab}
\centering
\begin{tabular}{ll}
\hline\hline
 Operator & \hspace{7mm} Eigenvalue \\
\hline~\\[-12pt]
\(\hat{H} = \beta - i\bm{\alpha}\cdot\bm{\nabla} + V(\rho)\)  & \hspace{7mm}\(\smash{E>0}\) \\[5pt]
\(\smash{\hat{p}_z = -\,i\frac{\partial}{\partial z}}\)  & \hspace{7mm}\(k\in\mathbb{R}\) \\[5pt]
\(\smash{\hat{J}_z = \tfrac{1}{2}\Sigma_z - i\frac{\partial}{\partial \phi}}\)  & \hspace{7mm}\(\smash{m_j=\pm1/2,\sm\pm3/2,\sm\dots}\) \\[5pt]
\(\hat{R}_z = \beta\Sigma_z + \gamma_5\hat{p}_z\)  & \hspace{7mm}\(\smash{\mu = \pm\sqrt{1+k^2}}\) \\[1mm]
\hline\hline
\end{tabular}
\end{table}

Inserting \eqref{Phi} into the TIDE reduces it to a coupled system of two ODEs for \(F\) and \(G\). Introducing 
\begin{equation}
    f(\rho) = \sqrt{\rho}\,F(\rho),\qquad \text{and}\qquad g(\rho)=-\,i\sqrt{\rho}\,G(\rho),
\end{equation}
renders this system real-valued:
\begin{equation}\label{CCC}
    \left(\frac{m_j}{\rho}\pm\frac{d}{d\rho}\right)\mqty{g\\f}=\Big(E\mp \mu-V(\rho)\Big)\,\mqty{f\\[-3pt]g}.
\end{equation}
We may therefore seek \(f,\sm g\) as real-valued functions.

This construction applies to arbitrary $V(\rho)$. For the specific potential of interest, \emph{viz.}, \(\smash{V=-\,v_0/\rho}\), one recognizes by direct inspection that the system \eqref{CCC} is identical to the radial equations arising from the spherical Coulomb--Dirac problem \cite[Sec.~9.6, Eq.~(1)]{GreinerWaveEquations}, with \(-\mu\) replacing the rest-mass energy \(mc^2\), \(m_j\) replacing the \(\kappa\) quantum number, and \(v_0\) replacing \(Z\alpha\). This system of equations has been extensively discussed in the literature; see \cite{hydrogen} for a historical survey.

Standard asymptotic analysis of that problem yields a power law \(\smash{\sim\rho^\gamma}\) at small \(\rho\) and exponential decay \(\smash{\sim e^{-\,\lambda\rho}}\) at large \(\rho\) for radially bound states, with \cite[Sec.~9.6, Eqs.~(5) and (6)]{GreinerWaveEquations}
\begin{equation}\label{gamma}
   \lambda = \sqrt{\mu^2-E^2},\qquad \text{and}\qquad \gamma = \sqrt{m_j^2-v_0^2}
\end{equation}
---real provided \(\smash{E<|\mu|}\) and \(\smash{|v_0|<\min|m_j|=1/2}\)---precisely the condition under which \(\hat{H}\) admits a distinguished self-adjoint realization \cite{Gallone,DES}. Combining both asymptotics, we seek solutions of the simple form
\begin{equation}
    f=A\sm \rho^{\gamma}e^{-\lambda\sm\rho},\qquad g=B\sm \rho^{\gamma}e^{-\lambda\sm\rho},
\end{equation}
with \(A\) and \(B\) being undetermined coefficients.

Inserting these Ans\"atze into Eqs.~\eqref{CCC} and equating coefficients of various powers of \(\rho\) to zero yields the constraints
\begin{subequations}
    \begin{align}
    &\big(\gamma+m_j\big)\sm B-v_0\sm A=0, &&\big(\gamma-m_j\big)\sm A+v_0\sm B=0, \label{AB}\\
    &(\mu+E)\sm B-\lambda A=0, &&(\mu-E)\sm A-\lambda B=0.\label{AABB}
    \end{align}
\end{subequations}
Both systems are degenerate---the determinants of Eqs.~\eqref{AB} and \eqref{AABB} vanish by virtue of \eqref{gamma}---leaving the ratio \(A/B\) fixed (rather than \(A\) and \(B\) individually). Specifically,
\begin{equation}\label{preSommerfeld}
    \frac{A}{B} = \frac{\gamma+m_j}{v_0}=\frac{\lambda}{\mu-E}.
\end{equation}
Since the TIDE is homogeneous in \(\Phi\), we set \(\smash{B = 1}\) hereafter, without loss of generality. 

Together with \eqref{gamma}, \eqref{preSommerfeld} determines
\begin{equation}
    E = \gamma \mu/m_j.
\end{equation}
Positive energy requires \(\smash{\text{sgn}(m_j) = \text{sgn}(\mu)}\)  (since \(\smash{\gamma>0}\)), fixing the sign of the \(\hat{R}_z\) eigenvalue to match that of \(\hat{J}_z\) [see Table \ref{EigenTab}]. In particular,
\begin{equation}
    E-1=\frac{\gamma}{|m_j|}\sqrt{1+k^2}-1 = \frac{k^2}{2}-\frac{v_0^2}{2\sm m_j^2}+\mathcal{O}\big(k^2v_0^2\big),
\end{equation}
for \(\smash{|k|,\sm v_0\ll1}\), thereby reproducing the correct nonrelativistic energy levels of the \(-\,v_0/\rho\) potential \cite{2dhydrogen}.

The stationary states thus read
\begin{equation}\label{exp2}
    \Phi:=\braket{\vb{r}\sm}{\sm m_j,k}=\rho^{\gamma-1/2}\mqty(\displaystyle\frac{\gamma+m_j}{v_0}\sm e^{i(m_j-1/2)\sm\phi}\\[10pt]\displaystyle\frac{\text{sgn}(m_j)\sm\sqrt{1+k^2}-1}{k} \,i\sm e^{i(m_j+1/2)\sm\phi}\\[10pt]\displaystyle\frac{\text{sgn}(m_j)\sm\sqrt{1+k^2}-1}{k} \left(\!\frac{\gamma+m_j}{v_0}\!\right)e^{i(m_j-1/2)\sm\phi}\\[10pt]\displaystyle i\sm e^{i(m_j+1/2)\sm\phi})\exp{\,ikz-\frac{\sqrt{1+k^2}}{|m_j|}\sm\big(\rho\sm v_0+i\gamma t\big)\sm},
\end{equation}
in the position representation.

In what follows, we consider only odd-parity wave functions, those fulfilling \(\beta\Phi(-\sm\vb{r})=-\sm\Phi(\vb{r})\) \footnote{In cylindrical coordinates, \(\smash{\vb{r}\mapsto-\,\vb{r}}\) is implemented via \((\rho,\phi,\allowbreak z) \mapsto(\rho,\phi+\pi,-\,z)\).}, formed by superposing the modes \(\ket{\pm1/2,k}\), given by
\begin{equation}
    \ket{\uparrow,k}=\frac{1}{2\sm i}\ket{\frac{1}{2},k}-\frac{1}{2\sm i}\ket{\frac{1}{2},-\,k},\qquad \ket{\downarrow,k}=\frac{1}{2}\ket{-\,\frac{1}{2},k}+\frac{1}{2}\ket{-\,\frac{1}{2},-\,k}.
\end{equation}
In terms of the functions \(\smash{\omega_\pm(k)=k^{-1}\sqrt{\sqrt{k^2+1}\pm 1}}\) and the single parameter \(\smash{\zeta=\tan^{-1}(v_0/\gamma)\overset{\eqref{gamma}}{=}\sin^{-1}(2\sm v_0)}\)---equivalently, \(v_0=\sin(\zeta)/2\) and \(\gamma=\cos(\zeta)/2\), we have
\begin{subequations}
    \begin{align}
        \braket{\vb{r}\sm}{\uparrow,k} &=  \rho^{(\cos\zeta\,-\,1)/2}\sm\mqty(\displaystyle\cot\frac{\zeta}{2}\sm\sin(kz)\\[8pt] \displaystyle k\sm\omega_-^2(k)\cos(kz)\sm e^{i\phi}\\[7pt]\displaystyle-\,i k\sm\omega_-^2(k)\cot\frac{\zeta}{2}\cos(kz)\\[8pt]\displaystyle i\sin(kz)\sm e^{i\phi})\exp{-\,(\rho\sm \sin\zeta+it\cos\zeta)\sm\sqrt{1+k^2}\,},\label{preup} \\
        \braket{\vb{r}\sm}{\downarrow,k} &=  \rho^{(\cos\zeta\,-\,1)/2}\sm\mqty(\displaystyle-\,\tan\frac{\zeta}{2}\sm\cos(kz)\sm e^{-\,i\phi}\\[8pt]\displaystyle k\sm\omega_+^2(k)\sin(kz)\\[7pt]\displaystyle i k\sm\omega_+^2(k)\tan\frac{\zeta}{2}\sm\sin(kz)\sm e^{-\,i\phi}\\[8pt]\displaystyle i\cos(kz))\exp{-\,(\rho\sm \sin\zeta+it\cos\zeta)\sm\sqrt{1+k^2}\,}.\label{predown}
    \end{align}
\end{subequations}
A short calculation shows that \(\smash{\braket{\uparrow,k}{\downarrow,k^\prime}=0}\). Also note that the states \(\ket{s,k}\) and \(\ket{s,-\,k}\), \(\smash{s=\uparrow,\downarrow}\) are linearly dependent; in particular, \(\smash{\ket{\uparrow,-\,k}=-\,\ket{\uparrow,k}}\) and \(\smash{\ket{\downarrow,-\,k}=\ket{\downarrow,k}}\). So, it suffices to consider \(\smash{k>0}\).

The states \(\ket{\uparrow,k}\) and \(\ket{\downarrow,k}\) as defined above have incommensurate norms:
\begin{equation}\label{IP}
    \braket{s,k}{s,k^\prime}=\delta\big(k-k^\prime\big)\,\Lambda(\zeta)\sm\big(1+k^2\big)^{-\sm\cos\zeta/2}\!\begin{cases}\,\displaystyle\omega_-^2(k)\csc^2(\zeta/2), &s=\uparrow\\\,\displaystyle \omega_+^2(k)\sec^2(\zeta/2), &s=\downarrow\end{cases}
\end{equation}
for \(\smash{k,\sm k^\prime>0}\) and \(\smash{s=\uparrow,\sm\downarrow}\). Here, \(\smash{\Lambda(\zeta)=(2\sm\pi)^2\Gamma(1+\cos\zeta)\sm(2\sin\zeta)^{-\sm 1-\cos\zeta}}\). In view of this, we rescale the kets to equalize their norms, letting
\begin{equation}
    \ket{\uparrow,k}\mapsto\frac{\sin(\zeta/2)}{\omega_-(k)}\ket{\uparrow,k}\kern-0.1em,\qquad \ket{\downarrow,k}\mapsto\frac{\cos(\zeta/2)}{\omega_+(k)}\ket{\downarrow,k}\kern-0.1em.
\end{equation}
Using the identity
\begin{equation}\label{iden}
    \frac{\sqrt{1+k^2}\pm1}{k\sm \omega_\pm(k)}=k\sm\omega_\pm=\frac{1}{\omega_\mp(k)},
\end{equation}
the rescaled kets in the position representation take the form \eqref{omegakets} of the main text.
\subsection{Evaluating \texorpdfstring{$I_\pm$}{~}}\label{AppIpm}
Here, we detail the exact evaluation of the longitudinal momentum integrals \eqref{IPM}. To obtain \eqref{IpmGold}, we incorporate \(a(k)\), Eq.~\eqref{simp}, into \eqref{IPM}, setting \(\smash{\widetilde{w} = w + w_0}\) and introducing \(\smash{f_+(x) = x}\) and \(\smash{f_-(x) = 1}\) for brevity:
\begin{align*}
    &\sqrt{\frac{\pi}{2}}\,I_{\pm}(z,w)=N\!\int_0^{\infty}\!\!\!dk~\cfrac{k\sm e^{-\,\widetilde{w}\sqrt{1+k^2}}}{\omega_\pm(k)\sqrt{1+k^2}}~\mqty{\displaystyle\cos\\[-3pt]\displaystyle\sin}\sm(kz)=\pm\,N\,\frac{\partial}{\partial z}\int_0^{\infty}\!\!\!dk~\cfrac{e^{-\,\widetilde{w}\sqrt{1+k^2}}}{\omega_\pm(k)\sqrt{1+k^2}}~\mqty{\displaystyle\sin\\[-3pt]\displaystyle\cos}\sm(kz)\\
    &\quad\overset{\eqref{iden}}{=}\pm\,N\,\frac{\partial}{\partial z}\int_0^{\infty}\!\!\!dk\,f_\pm(k)\, \omega_{-}^{\pm1}\sm\frac{e^{-\,\widetilde{w}\sqrt{1+k^2}}}{\sqrt{1+k^2}}\,\mqty{\displaystyle\sin\\[-3pt]\displaystyle\cos}\sm(kz)=\pm\,N\sqrt{\frac{\pi}{2}}\,\frac{\partial}{\partial z}\,f_\pm(z)\left[\sqrt{\widetilde{w}^2+z^2} + \widetilde{w}\right]^{\mp\sm 1/2}\frac{\exp(-\sqrt{\widetilde{w}^2+z^2}\,)}{\sqrt{\widetilde{w}^2+z^2}},
\end{align*}
via \cite[Sec.~3.962, formulae 1 and 2]{GH}, which directly yields \eqref{IpmGold}.

For the exponential weight \eqref{c2}, expressing the trigonometric functions in \eqref{IPM} as complex exponentials allows us to construct the integrals \(I_\pm\) via the auxiliary generating functions:
\begin{equation}
    J_\pm(z,w) = \sqrt{\frac{2}{\pi}}\int_0^{\infty}\!\!\frac{dk}{\omega_\pm(k)}\,\exp(-\,z\sm k-w\sqrt{1+k^2}\,),
\end{equation}
defined for complex \(w\) and \(z\) with positive real parts. In order to evaluate \(J_\pm\), we reformulate it as follows:
\begin{equation}\label{Kpmdef}
    J_\pm(z,w) = -\,\frac{\partial K_\pm}{\partial w},\quad \text{where}\quad K_\pm(z,w) = \sqrt{\frac{2}{\pi}}\int_0^{\infty}\!\frac{dk}{\omega_\pm(k)\sqrt{1+k^2}}\,\exp(-\,z\sm k-w\sqrt{1+k^2}\,),
\end{equation}
invoking the Leibniz integral rule, valid here since the integrand is analytic and integrably bounded with respect to \(w\).

By applying the identity \eqref{iden}, \(K_-\) reduces to a standard Laplace transform \cite[Sec.~2.2.5, formula 2]{Prudnikov} that can be expressed exactly in terms of the complementary error function (\(\text{erfc}\)). To evaluate \(K_+\), we apply Leibniz's rule a second time---now with respect to the parameter \(z\)---yielding:
\begin{align}
    K_+(z,w) = -\,\sqrt{\frac{2}{\pi}}\,\frac{\partial}{\partial z}\int_0^{\infty}\!\!\frac{dk}{k\sm\omega_+(k)\sqrt{1+k^2}}\,\exp(-\,z\sm k-w\sqrt{1+k^2}\,) = -\,\sqrt{\pi}\,e^{w}\frac{\partial}{\partial z}\sm\Big[ \text{erfc}\big(\eta_+\big)\sm \text{erfc}\big(\eta_-\big)\Big],
\end{align}
where \(\smash{\eta_\pm = \sqrt{w\pm\sqrt{w^2-z^2}}}\); see \cite[Sec.~2.2.5, formula 3]{Prudnikov}. Finally, evaluating the \(z\)-derivative and applying \eqref{Kpmdef} yields \(J_+\), which is structurally analogous to \(J_-\) and presented together in \eqref{Jpm}.
\end{document}